\documentclass
[superscriptaddress,secnumarabic,amssymb,amsmath,nobibnotes,aps,prd,showkeys,showpacs,nofootinbib,onecolumn,12pt]{revtex4}%
\usepackage{graphicx}
\usepackage{epsf}
\usepackage{bm}
\usepackage{amsmath}
\usepackage{amsfonts}
\usepackage{amssymb}%
\usepackage{pdflscape}
\providecommand{\U}[1]{\protect\rule{.1in}{.1in}}

\newcommand{\be}{\begin{equation}}
\newcommand{\ee}{\end{equation}}

\newcommand{\mincir}{\raise
-3.truept\hbox{\rlap{\hbox{$\sim$}}\raise4.truept\hbox{$<$}\ }}
\newcommand{\magcir}{\raise
-3.truept\hbox{\rlap{\hbox{$\sim$}}\raise4.truept\hbox{$>$}\ }}

\begin{document}
\title{Hyperbolic inflationary model with nonzero curvature}
\author{Andronikos Paliathanasis}
\email{anpaliat@phys.uoa.gr}
\affiliation{Institute of Systems Science, Durban University of Technology, PO Box 1334,
Durban 4000, South Africa}
\affiliation{Instituto de Ciencias F\'{\i}sicas y Matem\'{a}ticas, Universidad Austral de
Chile, Valdivia 5090000, Chile}
\author{Genly Leon}
\email{genly.leon@ucn.cl}
\affiliation{Departamento de Matem\'aticas, Universidad Cat\'olica del Norte, Avda. Angamos
0610, Casilla 1280 Antofagasta, Chile}
\affiliation{Institute of Systems Science, Durban University of Technology, PO Box 1334,
Durban 4000, South Africa}

\begin{abstract}
We consider a cosmological model consisting of two scalar fields defined in
the hyperbolic plane known as hyperbolic inflation. For the background space,
we consider a homogeneous and isotropic spacetime with nonzero curvature. We
study the asymptotic behaviour of solutions and we search for attractors in
the expanding regime. We prove that two hyperbolic inflationary stages are stable solutions that can solve the flatness problem and describe acceleration for both open and closed models, and additionally we obtain a Milne-like attractor solution for the open model. We also investigate the contracting branch obtaining mirror solutions with the opposite dynamical behaviours. 
\end{abstract}
\keywords{Scalar field; Multi-scalar field; Dynamical analysis; curvature}
\pacs{98.80.-k, 95.35.+d, 95.36.+x}
\date{\today}
\maketitle

\section{Introduction}

\label{sec1}

A simple mechanism that has been proposed to solve the homogeneity, isotropy, and flatness problems are the so-called cosmic inflation \cite{guth, lin}. In
the\ inflation, the universe has gone through a rapid expansion which has been
driven\ by an exotic matter source with a negative pressure component known as
inflation. The importance of rapid expansion is that the size of the universe
increases so fast that it loses its memory of the initial conditions. The
source and the nature of the inflationary mechanism are still unknown. There
are various approaches which are based on scalar fields
\cite{in01,in02,in03,in04,in05,bsjd}, on Chaplygin Gas \cite{in06,in07,in08}
or on the modification of the Einstein-Hilbert Action Integral
\cite{in09,in10,in11,in12}.

An inflationary model which has drawn the attention of cosmologists in recent
years is the so-called hyperbolic inflation \cite{ch5}. It is a two-scalar
field inflationary model in which the kinetic part of the scalar fields lies
on a hyperbolic plane. The model is inspired by the $\sigma$-theory and this
kind of model has been widely studied as a dark energy alternative in the past
\cite{sch1,sch2,sch3,ch3,ch4}. In hyperbolic inflation, for the exponential
potential, the inflationary era is attributed to a scaling attractor in which
the inflation consists of the two-scalar fields and it does not slow-roll
\cite{ch2}. Furthermore, because the scalar fields do not need to have the
same values at the beginning and the end of inflation, that means that there
are no observable non-Gaussianities in the power spectrum \cite{ch6,ch7}. The
analytic solution for the cosmological field equations of the hyperbolic
inflation in a spatially flat Friedmann--Lema\^{\i}tre--Robertson--Walker
(FLRW) background space was determined in \cite{sco01}. Other extensions were
proposed recently in \cite{sco02,sco03}. Recently, the scalar hyperbolic
inflationary model was investigated in the case of anisotropic spacetimes, in
which analytic solutions were determined \cite{sco03}, while the dynamics were
investigated in \cite{sco04}.

In this study, we investigate the dynamics of the hyperbolic inflation model
in the case of FLRW spacetime with nonzero curvature. Specifically, we are
interested to investigate if the given model can drive the dynamics so that a
future attractor be the inflationary solution or another spatially flat\ FLRW
universe. With this analysis, we shall understand further if this specific
multi-field model can solve the flatness problem. The dynamical analysis is an
essential approach for the study of physical viability for given gravitational
theories. See for instance \cite{dn1,dn2,dn3} and references therein. 

The plan of the paper is as follows. In Section \ref{sec2} we present the model of our consideration and we define
the field equations. Section \ref{sec3} includes the main analysis of this
study in which we investigate the asymptotic dynamics for the field equations.
Finally, in Section \ref{sec4} we discuss our results.

\section{Hyperbolic inflation}

\label{sec2}

We consider the two-scalar field model \cite{ch5}%
\begin{equation}
S=\int\sqrt{-g}dx^{4}\left(  R-\frac{1}{2}g^{\mu\nu}\nabla_{\mu}\phi
\nabla_{\nu}\phi-\frac{1}{2}g^{\mu\nu}e^{2\kappa\phi}\nabla_{\mu}\psi
\nabla_{\nu}\psi-V\left(  \phi\right)  \right)  , \label{sp.01}%
\end{equation}
where the two scalar fields $\phi\left(  x^{\mu}\right)  $ and $\psi\left(
x^{\nu}\right)  $ have kinetic terms which lie on a two-dimensional hyperbolic manifold.

For the background space we assume the FLRW universe%
\begin{equation}
ds^{2}=-dt^{2}+a^{2}\left(  t\right)  \left(  \frac{dr^{2}}{1-Kr^{2}}%
+r^{2}\left(  d\theta^{2}+\sin^{2}\theta d\phi^{2}\right)  \right)  ,
\label{sp.02}%
\end{equation}
where $K$ is the spatial curvature for the three-dimensional hypersurface. For
$K=0$, we have a spatially flat universe, for $K=1$ we have a closed universe
and for $K=-1$ the line element (\ref{sp.02}) describes an open universe. In
previous studies the cosmological model with Action Integral (\ref{sp.01}) has
been investigated for $K=0$ \cite{ch2}.
In the following we consider the case $K\neq0$. Moreover, we assume that the
scalar fields inherit the symmetries of the spacetime (\ref{sp.02}) which
means $\phi\left(  x^{\mu}\right)  =\phi\left(  t\right)  $ and $\psi\left(
x^{\mu}\right)  =\psi\left(  t\right)  $.

For the line element (\ref{sp.02}) and the Action Integral (\ref{sp.01}) we
end with the gravitational field equations
\begin{align}
& -3H^{2}+\frac{1}{2}\dot{\phi}^{2}+\frac{1}{2}e^{2\kappa\phi}\dot{\psi}%
^{2}+V\left(  \phi\right)  -3Ka^{-2}=0~, \label{sp.04}%
\\
& 2\dot{H}+3H^{2}+\frac{1}{2}\dot{\phi}^{2}+\frac{1}{2}e^{2\kappa\phi}\dot{\psi
}^{2}-V\left(  \phi\right)  + K a^{-2}=0~, \label{sp.05}%
\\
&  \ddot{\phi}+3H\dot{\phi}  -\kappa e^{2\kappa\phi}\dot{\psi}%
^{2}+V_{,\phi}\left(  \phi\right)  =0, \label{sp.06}%
 \\
&\ddot{\psi}+3H\dot{\psi}+2\kappa\dot{\phi}\dot{\psi}=0~, \label{sp.07}%
\end{align}
where $H=\frac{\dot{a}}{a}$ is the Hubble function.

We can define the effective energy density, $\rho_{eff}$, ~and the\ effective
pressure, $p_{eff},~$as
\begin{align}
& \rho_{eff}=\frac{1}{2}\dot{\phi}^{2}+\frac{1}{2}e^{2\kappa\phi}\dot{\psi}%
^{2}+V\left(  \phi\right)  , \label{sp.08}%
\\
& 
p_{eff}=\frac{1}{2}\dot{\phi}^{2}+\frac{1}{2}e^{2\kappa\phi}\dot{\psi}%
^{2}-V\left(  \phi\right)  ~. \label{sp.09}%
\end{align}
Thus, the parameter for the equation of state for the effective cosmological
fluid is
\begin{equation}
w_{eff}=\frac{p_{eff}}{\rho_{eff}}=\frac{\frac{1}{2}\dot{\phi}^{2}+\frac{1}%
{2}e^{2\kappa\phi}\dot{\psi}^{2}-V\left(  \phi\right)  }{\frac{1}{2}\dot{\phi
}^{2}+\frac{1}{2}e^{2\kappa\phi}\dot{\psi}^{2}+V\left(  \phi\right)  }~,
\label{sp.10}%
\end{equation}
while the deceleration parameter in the presence of spatial curvature is
$q=-1-\dot{H}/H^2$ and
$\Omega_{K}=K\left(  aH\right)  ^{-2}$.

\section{Dynamical analysis}

\label{sec3}

In order to investigate the dynamics and the asymptotic behaviour of the field
equations we define new dimensionless variables. Indeed, we assume the new
independent variable, $d\tau=\sqrt{H^{2}+|K| a^{-2}}dt$, and the new
dependent variables%
\begin{align}
&  \lambda=\frac{V'\left(\phi\right)}{V\left(\phi\right)}, x=\frac{\dot{\phi}}{\sqrt{6}\sqrt{H^{2}+|K|a^{-2}}},  y^{2}=\frac{V\left(  \phi\right)  }{3\left(  H^{2}+|K|a^{-2}\right)  },  \\ & z=e^{\kappa\phi}\frac{\dot{\psi}}{\sqrt{6}\sqrt{H^{2}+|K|a^{-2}}}, 
\eta=\frac{H}{\sqrt{H^{2}+|K|a^{-2}}}.
\end{align}
The functional forms of the field equations depend upon the sign of curvature $K$. That is, 
\begin{align}
 \frac{d\lambda}{d\tau}&=\sqrt{6}x\lambda^{2}\left(  \Gamma\left(
\lambda\right)  -1\right), \quad \Gamma\left(  \lambda\right)  =\frac{1}{\lambda
}\frac{V_{\phi\phi}}{V_{,\phi}}, \label{sp.16}\\
\frac{dx}{d\tau}& =    \frac{1}{2} \Big(\eta  x \left(\text{sgn}(K)+3 x^2-3
   y^2+3 z^2-4\right)  -\eta ^3 x (\text{sgn}(K)-1)   +\sqrt{6}
   \left(2 \kappa  z^2-\lambda  y^2\right)\Big), \label{sp.16B}\\
 \frac{dy}{d\tau}&=  \frac{y}{2|K|}  \Big(|
   K|  \left(\eta ^3+\eta  \left(3 x^2-3 y^2+3
   z^2+2\right)+\sqrt{6} \lambda  x\right)   +\eta\left(1 -\eta
   ^2\right) K\Big), \label{sp.16C}\\
\frac{dz}{d\tau}& =  \frac{1}{2} z \Big(-\left(\eta
   ^3 (\text{sgn}(K)-1)\right)-2 \sqrt{6} \kappa 
   x   +\eta  \left(\text{sgn}(K)+3 x^2-3
   y^2+3 z^2-4\right)\Big), \label{sp.16D}\\
\frac{d\eta}{d\tau}& =  \frac{\left(\eta ^2-1\right) \left(| K|  \left(\eta
   ^2+3 \left(x^2-y^2+z^2\right)\right)+(1-\eta ^2)K\right)}{2
   | K| }, \label{sp.16E}
\end{align}
with the Friedman constraint 
\begin{equation}
    \eta ^2-\left(\eta ^2-1\right) \text{sgn}(K)-x^2-y^2-z^2=0. \label{EQ17}
\end{equation}
For the scalar field potential we assume the exponential function, $V\left(
\phi\right)  =V_{0}e^{\lambda\phi}$ \cite{ch5}. In this case, $\lambda$ is a
constant parameter. Hence, we omit the additional equation 
\eqref{sp.16} that is trivially satisfied. 

The deceleration parameter can be expressed in terms of the normalized variables as 
\begin{align}
    q& =\frac{\eta ^2-\eta ^2 \text{sgn}(K)+\text{sgn}(K)+3 x^2-3 y^2+3 z^2}{2 \eta ^2} \nonumber \\
    & =\left(1-\frac{1}{\eta ^2}\right)
   \text{sgn}(K)+\frac{3 \left(x^2+z^2\right)}{\eta ^2}-1,
\end{align}
and the effective equation of state parameter as 
\begin{align}
w_{eff}=1-\frac{2 y^2}{x^2+y^2+z^2}= \frac{2 \left(x^2+z^2\right)}{\eta ^2+(1-\eta ^2)\text{sgn}(K)}-1.
\end{align}
Now, we investigate separately the two cases, $K=1$ and $K=-1$. 

\subsection{Closed Universe}

For $K=1$ the Friedmann constraint \eqref{EQ17} reduces to%
\begin{equation}
1-x^{2}-y^{2}-z^{2}=0. \label{sp. 11}%
\end{equation}
We determine the stationary points of the system \eqref{sp.16}, \eqref{sp.16B}, \eqref{sp.16C}, \eqref{sp.16D} and 
\eqref{sp.16E} for $K=1$ and we study their
stability properties. Each stationary point corresponds to a specific epoch of
the cosmological evolution. It is important to mention that from
(\ref{sp. 11}) parameters $x,y,z$ are constraints in a unitary sphere.
Moreover, we can use the constraint equation (\ref{sp. 11}) to reduce the
dimension of the dynamical system. Indeed, we assume $y=\sqrt{1-x^{2}-z^{2}}$.
We remark that $y\rightarrow-y$ is a discrete symmetry for the field
equations. In addition, by definition parameter $\eta$ is constrained as
$\left\vert \eta\right\vert \leq1$. The case $\left\vert \eta\right\vert =1$
corresponds to spatially flat universes. A positive value of $\eta$ denotes an
expanding universe, i.e. $H>0$. \ Thus, in the following we focus on the
expansion era \ that is we assume $\eta>0$.
\begin{table*}[]
\centering
\resizebox{\textwidth}{!}{
\begin{tabular}{|c|c|c|c|c|c|c|} \hline 
Label &   $x$& $z$ & $\eta$ & $e_1(P)$ & $e_2(P)$ & $e_3(P)$  \\\hline
 $A_1^{+}$& $ 1$ & $0$ & $ 1$ & $4$ & $-\sqrt{6} \kappa$  & $6+\sqrt{6} \lambda$ \\\hline
 $A_1^{-}$ & $-1$ & $0$ & $ 1$ & $4$ & $\sqrt{6} \kappa$  &    $6-\sqrt{6} \lambda$  \\\hline
 $\bar{A}_1^{+}$& $1$ &$ 0$ & $-1$ & $-4$ & $-\sqrt{6} \kappa$  &    $-(6-\sqrt{6} \lambda)$ \\\hline 
 $\bar{A}_1^{-}$     &    $-1$ & $0$ & $-1$ & $-4$ & $\sqrt{6} \kappa$  & $-(6+\sqrt{6} \lambda)$ \\\hline 
 $A_2$ & $ -\frac{\lambda }{\sqrt{6}}$ & $ 0$ & $1$ &   $ \frac{1}{2} \left(\lambda ^2-6\right)$ & $\lambda ^2-2$ & $-3+\frac{1}{2} \lambda (2 \kappa +\lambda ) $ \\\hline 
 $\bar{A}_2$ & $ \frac{\lambda }{\sqrt{6}}$ & $0$ & $-1$ & $ -\frac{1}{2} \left(\lambda ^2-6\right)$ & $-(\lambda ^2-2)$ & $3-\frac{1}{2} \lambda (2 \kappa +\lambda ) $\\ \hline
 $A_3^{+}$& $ -\frac{\sqrt{6}}{2 \kappa +\lambda } $ & $\frac{\sqrt{2
   \kappa  \lambda +\lambda ^2-6}}{\sqrt{(2 \kappa +\lambda
   )^2}}$& $1$ & $4-\frac{12 \kappa }{2 \kappa +\lambda }$
   & $-\frac{3 \kappa +i   \sqrt{3 \kappa \left[8 \kappa ^2
   \lambda +\kappa  \left(8 \lambda ^2-27\right)+2 \lambda 
   \left(\lambda ^2-6\right)\right]}}{2 \kappa +\lambda }$ & $-\frac{3
   \kappa -i   \sqrt{3 \kappa \left[8 \kappa ^2
   \lambda +\kappa  \left(8 \lambda ^2-27\right)+2 \lambda 
   \left(\lambda ^2-6\right)\right]}}{2 \kappa +\lambda }$ \\\hline 
 $A_3^-$  & $-\frac{\sqrt{6}}{2 \kappa +\lambda }$ & $-\frac{\sqrt{2
   \kappa  \lambda +\lambda ^2-6}}{\sqrt{(2 \kappa +\lambda
   )^2}}$ & $1$ & $ 4-\frac{12 \kappa }{2 \kappa +\lambda }$
   & $-\frac{3 \kappa +i   \sqrt{3 \kappa \left[8 \kappa ^2
   \lambda +\kappa  \left(8 \lambda ^2-27\right)+2 \lambda 
   \left(\lambda ^2-6\right)\right]}}{2 \kappa +\lambda }$ & $-\frac{3
   \kappa -i   \sqrt{3 \kappa \left[8 \kappa ^2
   \lambda +\kappa  \left(8 \lambda ^2-27\right)+2 \lambda 
   \left(\lambda ^2-6\right)\right]}}{2 \kappa +\lambda }$ \\\hline
 $\bar{A}_3^+$ &  $\frac{\sqrt{6}}{2 \kappa +\lambda }$ & $\frac{\sqrt{2
   \kappa  \lambda +\lambda ^2-6}}{\sqrt{(2 \kappa +\lambda
   )^2}}$ & $ -1$ & $-4+\frac{12 \kappa }{2 \kappa +\lambda }$
   & $\frac{3 \kappa +i   \sqrt{3 \kappa \left[8 \kappa ^2
   \lambda +\kappa  \left(8 \lambda ^2-27\right)+2 \lambda 
   \left(\lambda ^2-6\right)\right]}}{2 \kappa +\lambda }$ & $\frac{3
   \kappa -i   \sqrt{3 \kappa \left[8 \kappa ^2
   \lambda +\kappa  \left(8 \lambda ^2-27\right)+2 \lambda 
   \left(\lambda ^2-6\right)\right]}}{2 \kappa +\lambda }$ \\\hline
  $\bar{A}_3^-$ & $\frac{\sqrt{6}}{2 \kappa +\lambda }$ & $-\frac{\sqrt{2
   \kappa  \lambda +\lambda ^2-6}}{\sqrt{(2 \kappa +\lambda
   )^2}}$ & $ -1$ & $-4+\frac{12 \kappa }{2 \kappa +\lambda }$
   & $\frac{3 \kappa +i   \sqrt{3 \kappa \left[8 \kappa ^2
   \lambda +\kappa  \left(8 \lambda ^2-27\right)+2 \lambda 
   \left(\lambda ^2-6\right)\right]}}{2 \kappa +\lambda }$ & $\frac{3
   \kappa -i  \sqrt{3 \kappa \left[8 \kappa ^2
   \lambda +\kappa  \left(8 \lambda ^2-27\right)+2 \lambda 
   \left(\lambda ^2-6\right)\right]}}{2 \kappa +\lambda }$ \\\hline
 $A_4^+$ & $-\frac{1}{\sqrt{3}}$ & $ 0$ & $\frac{\lambda}{\sqrt{2}}$ & $\sqrt{2} (\kappa -\lambda )$ & $-\frac{\lambda +\sqrt{8-3\lambda ^2}}{\sqrt{2}}$ & $-\frac{\lambda -\sqrt{8-3 \lambda ^2}}{\sqrt{2}}$\\\hline
 $A_4^{-}$&$ \frac{1}{\sqrt{3}}$ & $0$ & $-\frac{\lambda}{\sqrt{2}}$ & $-\sqrt{2} (\kappa -\lambda )$& $\frac{\lambda +\sqrt{8-3 \lambda   ^2}}{\sqrt{2}}$& $\frac{\lambda -\sqrt{8-3 \lambda ^2}}{\sqrt{2}}$ \\\hline
    \end{tabular}}
    \caption{The stationary points $A=\left(  x\left(  A\right)  ,z\left(  A\right)
,\eta\left(  A\right)  \right)$ and $\bar{A}=\left(  x\left(\bar{A}\right)  ,z\left(\bar{A}\right)
,\eta\left(\bar{A}\right)  \right)$ of the dynamical system \eqref{sp.12}, \eqref{sp.14}, \eqref{sp.15}.}
    \label{Tab1}
\end{table*}
Hence, we obtain the reduced system: 
\begin{align}
\frac{dx}{d\tau} & =3 \eta  x \left(x^2+z^2-1\right)+\sqrt{\frac{3}{2}}
   \left(\lambda  \left(x^2-1\right)+z^2 (2 \kappa +\lambda
   )\right), \label{sp.12}%
\\
\frac{dz}{d\tau} & =z \left(3 \eta  \left(x^2+z^2-1\right)-\sqrt{6}
   \kappa  x\right), \label{sp.14}%
\\
\frac{d\eta}{d\tau} & =\left(\eta ^2-1\right) \left(3 x^2+3
   z^2-1\right). \label{sp.15}%
\end{align}%
The stationary points $A=\left(  x\left(  A\right)  ,z\left(  A\right)
,\eta\left(  A\right)  \right)$ and $\bar{A}=\left(  x\left(\bar{A}\right)  ,z\left(\bar{A}\right)
,\eta\left(\bar{A}\right)  \right)$ of the dynamical system \eqref{sp.12}, \eqref{sp.14}, \eqref{sp.15} are represented in table \ref{Tab1}. 
Since the points $\bar{A}$ have the time-reversal dynamical behaviour of the related points $A$ under the change $(\tau, x, \eta) \mapsto (-\tau, -x,-\eta)$. $A_1^+$ maps onto $\bar{A}_1^-$, $A_1^-$ maps onto $\bar{A}_1^+$, $A_2$ maps onto $\bar{A}_2$,  $A_3^+$ maps onto $\bar{A}_3^-$, $A_3^-$ maps onto $\bar{A}_3^+$ and $A_4^-$ maps onto $A_4^+$ (where we have suppressed the bar). Then,  we focus on points $A$'s: 
\begin{align*}
& A_{1}^{\pm}=\left(  \pm1,0,1\right), A_{2}=\left( -\frac{\lambda}{\sqrt{6}%
},0,1\right),\\ 
& A_{3}^{\pm}=\left(-\frac{\sqrt{6}}{2\kappa+\lambda},\pm\sqrt{\frac
{\lambda^{2}+2\kappa\lambda-6}{\left(  2\kappa+\lambda\right)  ^{2}}%
},1\right),  A_{4}^{\pm}=\left(  \mp \frac{1}{\sqrt{3}},0,\pm\frac{\lambda}{\sqrt{2}%
}\right).     
\end{align*}
Stationary points, $A_{1}^{\pm}$, $A_{2}$ and $A_{3}^{\pm}$ describe
spatially flat universes. For the asymptotic solution at point $A_{1}^{\pm}$
we derive $w_{eff}\left(  A_{1}^{\pm}\right)  =1$ and $q\left(  A_{1}^{\pm
}\right)  =2$, which means that the universe is dominated by the kinetic part
of the scalar field $\phi\left(  t\right)  $. Point $A_{2}$ provides
$w_{eff}\left(  A_{2}\right)  =\frac{\lambda^{2}-3}{3}$,~$q\left(
A_{2}\right)  =\frac{\lambda^{2}-2}{2}$, which is nothing other than
the quintessence scaling solution of  \cite{dn1}. Acceleration occurs for
$\left\vert \lambda\right\vert <\sqrt{2}$. 

Points $A_{3}^{\pm}$ describe the hyperbolic inflationary solutions,
$w_{eff}\left(  A_{3}^{\pm}\right)  =1-\frac{4\kappa}{2\kappa+\lambda}$ ,
$q\left(  A_{3}^{\pm}\right)  =2-\frac{6\kappa}{2\kappa+\lambda}$.$\ $Points
are real and physically acceptable when $2\kappa+\lambda\neq0$ and
$\lambda^{2}+2\kappa\lambda-6>0$. Acceleration occurs when $\left\{
\lambda\leq-\sqrt{2},\kappa<\lambda\right\}  $, or~$\left\{  -\sqrt{2}<\lambda<0,
\kappa<\frac{6-\lambda^{2}}{2\lambda}\right\}  $, or $\left\{  0<\lambda<\sqrt
{2},\kappa>\frac{6-\lambda^{2}}{2\lambda}\right\}  $, or $\left\{  \lambda
\geq\sqrt{2},\kappa>\lambda\right\}  $.

Furthermore, points $A_{4}^{\pm}$ exist for $\lambda^2<2$ and describe Milne-like solutions with $a\left(
t\right)  =a_{0}t$ and $w_{eff}\left(  A_{4}^{\pm}\right)  =-\frac{1}{3}$,
$q\left(  A_{4}^{\pm}\right)  =0$. Point $A_{4}^{+}$ describes an expanding
universe for $\lambda>0$, while $A_{4}^{-}$ corresponds to an expanding
universe for $\lambda<0$.

To infer the stability properties of the stationary points we derive
the eigenvalues of the linearised system around the stationary points. Indeed,
for the points $A_{1}^{\pm}$ we find the eigenvalues $e_{1}\left(  A_{1}^{\pm
}\right)  =4$, $e_{2}\left(  A_{1}^{\pm}\right)  =\mp\sqrt{6}\kappa~$,
$e_{3}\left(  A_{1}^{\pm}\right)  =6\pm\sqrt{6}\lambda$. Thus, points $A_{1}^{\pm}$ are nonhyperbolic for $\kappa=0$ or $\lambda= \mp \sqrt{6}$. Point
$A_{1}^{+}$ is a source for $\lambda>-\sqrt{6}$ and $\kappa<0$. Point
$A_{1}^{-}$ is a source for $\lambda<\sqrt{6}$ and $\kappa>0$. Otherwise,
points $A_{1}^{\pm}$ are saddle points.

$A_2$ exists for $-\sqrt{6}\leq \lambda \leq \sqrt{6}$. We derive the eigenvalues $e_{1}\left(  A_{2}\right)
=-3+\frac{\lambda^{2}}{2}$, $e_{2}\left(  A_{2}\right)  =\lambda^{2}-2$ and
$e_{3}(A_2)=-3+\frac{1}{2}\lambda\left(  2\kappa+\lambda\right)  $. Thus, point $A_2$ is nonhyperbolic for $\lambda^2=2$ or $\lambda^2=6$ or $\lambda\left(2\kappa
+\lambda\right)=6$. Point
$A_{2}$ is a sink when $\lambda^{2}<2$ and $\lambda\left(2\kappa
+\lambda\right)  <6$, otherwise point $A_{2}$ is a saddle point. 

Points $A_{3}^{\pm}$ exist when $2\kappa+\lambda\neq0$ and
$\lambda^{2}+2\kappa\lambda-6>0$. The eigenvalues are $e_{1}\left(  A_{3}^{\pm}\right)  =4\left(
1-3\frac{\kappa}{2\kappa+\lambda}\right)$, $e_{2,3}\left(  A_{3}^{\pm
}\right)  =-\frac{3\kappa\pm i\sqrt{3\kappa\left(  8\kappa^{2}\lambda
+2\lambda\left(  \lambda^{2}-6\right)  +\kappa\left(  8\lambda^{2}-27\right)
\right)  }}{2\kappa+\lambda}$. We define $\kappa_{\pm}=\frac{27-8 \lambda ^2}{16 \lambda }\pm\frac{1}{16} \sqrt{3}
   \sqrt{-\frac{16 \lambda ^2-243}{\lambda ^2}}$. When hyperbolic, $A_{3}^{\pm}$  can
be sink for 
$\left\{-\sqrt{\frac{13}{6}}<\lambda \leq -\sqrt{2},  \kappa_-\leq \kappa <\lambda \right\}$, or  $ \left\{-\sqrt{2}<\lambda <0, 
 \kappa_-\leq \kappa
   <\frac{6-\lambda ^2}{2 \lambda }\right\}$, or $ \left\{0<\lambda <\sqrt{2},  \frac{6-\lambda ^2}{2 \lambda }<\kappa
   \leq \kappa_+\right\}$, or  $ \left\{\sqrt{2}\leq \lambda <\sqrt{\frac{13}{6}},  \lambda <\kappa \leq \kappa_+\right\}$, or $ \left\{\lambda \leq
   -\sqrt{\frac{13}{6}},  \kappa <\lambda \right\}$, or    $ \left\{-\sqrt{\frac{13}{6}}<\lambda <0,  \kappa <\kappa_-\right\}$, or  $
   \left\{0<\lambda <\sqrt{\frac{13}{6}},  \kappa > \kappa_+\right\}$, or $\left\{\lambda \geq \sqrt{\frac{13}{6}},  \kappa >\lambda
   \right\}$. They can be a saddle otherwise. 

Finally,  $A_{4}^{\pm}$  exists when $|\lambda|\leq \sqrt{2}$ with   eigenvalues 
$e_{1}\left(  A_{4}^{\pm}\right)  =\pm\sqrt{2}\left(\kappa - \lambda\right)
$, $e_{2}\left(  A_{4}^{\pm}\right)  =\mp\frac{\lambda-\sqrt{8- 3\lambda^{2}}%
}{\sqrt{2}}$ and $e_{3}\left(  A_{4}^{\pm}\right)  =\mp\frac{\lambda
+\sqrt{8  - 3\lambda^{2}}}{\sqrt{2}}$ from which we conclude that in the expanding
region the stationary points are saddle points.

We conclude that there is an attractor in the expanding branch for the field
equations in the presence of positive spatial curvature dominated by quintessence 
$A_{2}$,  that is a sink  when $\lambda^{2}<2$ and $\lambda\left(2\kappa
+\lambda\right)  <6$.
Moreover, the point $A_3^{\pm}$, that exist when $2\kappa+\lambda\neq0$ and
$\lambda^{2}+2\kappa\lambda-6>0$ and correspond to inflationary power-law solutions can be attractors. 
We continue by
considering the negative curvature.

\subsection{Open Universe}

For an open FLRW universe and $K=-1$ the constraint \eqref{EQ17} is  
\begin{equation}
 x^{2}+y^{2}+z^{2}= 2\eta^{2}%
-1.   \label{sp.26}
\end{equation}
With the use of the constraint equation (\ref{sp.26}) the reduced
dynamical system lies on the three-dimensional surface. By definition $\eta^{2}\leq1$. Consequently, parameters
$\left\{  x,y,z\right\}  $ are constrained , that is, $\frac{1}{2}\leq\eta^{2}\leq1$. To compare with the closed FLRW case, we assume $y=\sqrt{2 \eta^2 -1 -x^{2}-z^{2}}$.
We remark that $y\rightarrow-y$ is a discrete symmetry for the field
equations.
A positive value of $\eta$ denotes an
expanding universe, i.e. $H>0$.\ Thus in the following we focus on the
expansion era \ that is we assume $\eta>0$.

Hence, the field equations are%
\begin{align}
\frac{dx}{d\tau} & = -\sqrt{6} \eta ^2 \lambda +\eta  x \left(3 x^2+3
   z^2-1\right)   +\sqrt{\frac{3}{2}} \left(\lambda 
   \left(x^2+1\right)+z^2 (2 \kappa +\lambda )\right)-2 \eta ^3
   x, \label{sp.21}%
\\
\frac{dz}{d\tau} & = z \left(-2 \eta ^3+\eta  \left(3 x^2+3
   z^2-1\right)-\sqrt{6} \kappa  x\right), \label{sp.23}
\\
\frac{d\eta}{d\tau}& =\left(\eta ^2-1\right)
   \left(-2 \eta ^2+3 x^2+3 z^2+1\right). \label{sp.24}%
\end{align}%
\begin{table*}[]
\centering
\resizebox{\textwidth}{!}{
\begin{tabular}{|c|c|c|c|c|c|c|} \hline 
Label &   $x$& $z$ & $\eta$ & $e_1(P)$ & $e_2(P)$ & $e_3(P)$  \\\hline
$B_1^+$ & $ 1$ & $ 0$ & $ 1$ & $4$ & $-\sqrt{6} \kappa$  & $6+ \sqrt{6}
   \lambda$ \\\hline
$B_1^-$ & $ -1$ & $ 0$ & $ 1$&$ 4 $& $\sqrt{6} \kappa $ &
   $6-\sqrt{6} \lambda $ \\ \hline
$\bar{B}_1^+$ & $ 1$ & $ 0$ & $ -1$ & $-4$ & $-\sqrt{6} \kappa$  &
   $-\left(6-\sqrt{6} \lambda \right)$ \\\hline
$\bar{B}_1^-$   & $ -1$ & $ 0$ & $ -1$ & $-4$ & $\sqrt{6} \kappa$  &
  $-\left(6+ \sqrt{6}
   \lambda\right)$ \\\hline
$B_2$    & $ -\frac{\lambda }{\sqrt{6}}$ & $ 0 $& $ 1$ &
  $\frac{1}{2} \left(\lambda ^2-6\right)$ &$ \lambda ^2-2$ & $\kappa
    \lambda +\frac{\lambda ^2}{2}-3$ \\\hline
$\bar{B}_2$    & $\frac{\lambda }{\sqrt{6}}$ & $ 0 $& $-1$ &
  $-\frac{1}{2} \left(\lambda ^2-6\right)$ &$ -\left(\lambda ^2-2\right)$ & $-\left(\kappa
    \lambda +\frac{\lambda ^2}{2}-3\right)$ \\\hline
$B_3^+$ & $ -\frac{\sqrt{6}}{2 \kappa +\lambda }$ & $ \frac{\sqrt{2
   \kappa  \lambda +\lambda ^2-6}}{\sqrt{(2 \kappa +\lambda
   )^2}}$ &  $1$ & $4-\frac{12 \kappa }{2 \kappa +\lambda }$
      & $-\frac{3 \kappa + i  \sqrt{3 \kappa\left[8 \kappa
   ^2 \lambda +\kappa  \left(8 \lambda ^2-27\right)+2 \lambda 
   \left(\lambda ^2-6\right)\right]}}{2 \kappa +\lambda }$ &  $-\frac{3 \kappa - i  \sqrt{3 \kappa\left[8 \kappa
   ^2 \lambda +\kappa  \left(8 \lambda ^2-27\right)+2 \lambda 
   \left(\lambda ^2-6\right)\right]}}{2 \kappa +\lambda }$ 
\\\hline
$B_3^{-}$ & $ -\frac{\sqrt{6}}{2 \kappa +\lambda }$ & $ -\frac{\sqrt{2
   \kappa  \lambda +\lambda ^2-6}}{\sqrt{(2 \kappa +\lambda
   )^2}}$ & $ 1$ & $4-\frac{12 \kappa }{2 \kappa +\lambda }$
   & $-\frac{3 \kappa + i  \sqrt{3 \kappa\left[8 \kappa
   ^2 \lambda +\kappa  \left(8 \lambda ^2-27\right)+2 \lambda 
   \left(\lambda ^2-6\right)\right]}}{2 \kappa +\lambda }$ &
    $-\frac{3 \kappa - i  \sqrt{3 \kappa\left[8 \kappa
   ^2 \lambda +\kappa  \left(8 \lambda ^2-27\right)+2 \lambda 
   \left(\lambda ^2-6\right)\right]}}{2 \kappa +\lambda }$   \\ \hline
$\bar{B}_3^+$ & $ \frac{\sqrt{6}}{2 \kappa +\lambda }$ & $ \frac{\sqrt{2
   \kappa  \lambda +\lambda ^2-6}}{\sqrt{(2 \kappa +\lambda
   )^2}}$ & $ -1$ & $\frac{12 \kappa }{2 \kappa +\lambda }-4$
   & $\frac{3 \kappa + i  \sqrt{3 \kappa\left[8 \kappa
   ^2 \lambda +\kappa  \left(8 \lambda ^2-27\right)+2 \lambda 
   \left(\lambda ^2-6\right)\right]}}{2 \kappa +\lambda }$ &
    $\frac{3 \kappa - i  \sqrt{3 \kappa\left[8 \kappa
   ^2 \lambda +\kappa  \left(8 \lambda ^2-27\right)+2 \lambda 
   \left(\lambda ^2-6\right)\right]}}{2 \kappa +\lambda }$ \\ \hline
$\bar{B}_3^{-}$ & $\frac{\sqrt{6}}{2 \kappa +\lambda }$ & $ -\frac{\sqrt{2
   \kappa  \lambda +\lambda ^2-6}}{\sqrt{(2 \kappa +\lambda
   )^2}}$ &  $-1$ & $\frac{12 \kappa }{2 \kappa +\lambda }-4$
   & $\frac{3 \kappa + i  \sqrt{3 \kappa\left[8 \kappa
   ^2 \lambda +\kappa  \left(8 \lambda ^2-27\right)+2 \lambda 
   \left(\lambda ^2-6\right)\right]}}{2 \kappa +\lambda }$ &  $\frac{3 \kappa - i  \sqrt{3 \kappa\left[8 \kappa
   ^2 \lambda +\kappa  \left(8 \lambda ^2-27\right)+2 \lambda 
   \left(\lambda ^2-6\right)\right]}}{2 \kappa +\lambda }$  \\ \hline
$B_4^+$ & $ \frac{1}{\sqrt{3 \lambda ^2-3}}$ & $ 0$ & $
   -\sqrt{\frac{\lambda^2}{ 2 (\lambda ^2-1)}}$ &
   $\frac{\sqrt{2} (\lambda -\kappa )}{\sqrt{\lambda ^2-1}}$ &
   $\frac{\lambda + \sqrt{8-3 \lambda ^2}}{\sqrt{2}
   \sqrt{\lambda ^2-1}}$ & $\frac{\lambda -  \sqrt{8-3 \lambda
   ^2}}{\sqrt{2} \sqrt{\lambda ^2-1}}$ \\\hline
$B_4^{-}$ & $-\frac{1}{\sqrt{3 \lambda ^2-3}}$ & $0$ & $
  \sqrt{\frac{\lambda^2}{ 2 (\lambda ^2-1)}}$ &
   $-\frac{\sqrt{2} (\lambda  -\kappa)}{\sqrt{\lambda ^2-1}} $&
  $ -\frac{\lambda +\sqrt{8-3 \lambda ^2}}{\sqrt{2}
   \sqrt{\lambda ^2-1}}$&$ -\frac{\lambda - \sqrt{8- 3 \lambda
   ^2}}{\sqrt{2} \sqrt{\lambda ^2-1}}$ \\\hline
$B_5$ &$0$ & $0$ & $\frac{1}{\sqrt{2}}$ &$ -\sqrt{2}$ &
  $ -\sqrt{2}$ &$ \sqrt{2}$ \\\hline
$\bar{B}_5$ & $ 0$ & $0$  & $-\frac{1}{\sqrt{2}}$ & $\sqrt{2}$ &
  $ \sqrt{2}$ & $-\sqrt{2}$ \\\hline
    \end{tabular}}
    \caption{The stationary points $B=\left(  x\left(B\right)  ,z\left(B\right)
,\eta\left(B\right)  \right)$ and $\bar{B}=\left(  x\left(\bar{B}\right)  ,z\left(\bar{B}\right)
,\eta\left(\bar{B}\right)  \right)$ of the dynamical system \eqref{sp.21}, \eqref{sp.23}, \eqref{sp.24}.}
    \label{tab2}
\end{table*}
The stationary points $B=\left(  x\left(B\right)  ,z\left(B\right)
,\eta\left(B\right)  \right)$ and $\bar{B}=\left(  x\left(\bar{B}\right)  ,z\left(\bar{B}\right)
,\eta\left(\bar{B}\right)  \right)$ of the dynamical system \eqref{sp.21}, \eqref{sp.23}, \eqref{sp.24} are presented in 
table \ref{tab2}.
Since the points $\bar{B}$ have the time-reversal dynamical behaviour of the related points $B$ under the change $(\tau, x, \eta) \mapsto (-\tau, -x,-\eta)$. $B_1^+$ maps onto $\bar{B}_1^-$, $B_1^-$ maps onto $\bar{B}_1^+$, $B_2$ maps onto $\bar{B}_2$,  $B_3^+$ maps onto $\bar{B}_3^-$, $B_3^-$ maps onto $\bar{B}_3^+$, $B_4^-$ maps onto $B_4^+$ (where we have suppressed the bar), $B_5$ maps onto   $\bar{B}_5$. Then, we focus on points $B$'s:
\begin{align*}
& B_{1}^{\pm}=\left(  \pm1,0,1\right), B_{2}=\left(  -\frac{\lambda}{\sqrt{6}%
},0,1\right),   \\
& B_{3}^{\pm}=\left(  -\frac{\sqrt{6}}{2\kappa+\lambda}, \pm\sqrt{\frac{\lambda^{2}+2\kappa\lambda-6}{\left(
2\kappa+\lambda\right)  ^{2}}},1\right), 
\end{align*}
\begin{align*}
& B_{4}^{\pm}=\left(  \pm\frac{1}{\sqrt{3\left(  \lambda^{2}-1\right)  }}%
,0,  \mp\sqrt{\frac{\lambda^2}{ 2 (\lambda ^2-1)}}\right), B_{5}=\left(  0,0,\frac{1}{\sqrt{2}}\right).
\end{align*}
Stationary points $B_{1}^{\pm}$, $B_{2}$ and $B_{3}^{\pm}$ describe spatially
flat FLRW asymptotic solutions with physical properties similar with that of
points $A_{1}^{\pm},$~$A_{2}$ and $A_{3}^{\pm}$, respectively.

The eigenvalues
of the linearised system around the stationary points $B_{1}^{\pm}$ and
$B_{2}$ are derived to be $e_{1}\left(  B_{1}^{\pm}\right)  =4$,
$e_{2}\left(  B_{1}^{\pm}\right)  =\mp\sqrt{6}\kappa$ ,~$e_{3}\left(
B_{1}^{\pm}\right)  =6\pm \sqrt{6}
   \lambda$ and $e_{1}\left(
B_{2}\right)  =-3+\frac{\lambda^{2}}{2}$, $e_{2}\left(  B_{2}\right)
=-2+\lambda^{2}$ ,~$e_{3}\left(  B_{2}\right)  =-3+\frac{1}{2}\lambda\left(
2\kappa+\lambda\right)  $. Thus, points $B_{1}^{\pm}$ can be a source when
$\left\{  \pm\kappa<0, 6\pm \sqrt{6}
   \lambda>0\right\}$. 
$B_{2}$ is real only for
$\lambda^{2}<6$ and can be an attractor for $\lambda^{2}<2$ and $\lambda\left(
2\kappa+\lambda\right)<6$. Consequently this hyperbolic
inflationary solution is a future attractor. 

As far as points
$B_{3}^{\pm}$ are concerned, they exist when $2\kappa+\lambda\neq0$ and
$\lambda^{2}+2\kappa\lambda-6>0$, and the eigenvalues are
$e_1(B_{3}^{\pm})=4-\frac{12 \kappa }{2 \kappa +\lambda }$, $e_2(B_{3}^{\pm})=-\frac{3 \kappa + i  \sqrt{3 \kappa\left[8 \kappa
   ^2 \lambda +\kappa  \left(8 \lambda ^2-27\right)+2 \lambda 
   \left(\lambda ^2-6\right)\right]}}{2 \kappa +\lambda }$ and $e_3(B_{3}^{\pm})=-\frac{3 \kappa - i  \sqrt{3 \kappa\left[8 \kappa
   ^2 \lambda +\kappa  \left(8 \lambda ^2-27\right)+2 \lambda 
   \left(\lambda ^2-6\right)\right]}}{2 \kappa +\lambda }$. Hence, when hyperbolic, $B_{3}^{\pm}$  can
be sink for 
$\left\{-\sqrt{\frac{13}{6}}<\lambda \leq -\sqrt{2},  \kappa_-\leq \kappa <\lambda \right\}$, or  $ \left\{-\sqrt{2}<\lambda <0, 
 \kappa_-\leq \kappa
   <\frac{6-\lambda ^2}{2 \lambda }\right\}$, or $ \left\{0<\lambda <\sqrt{2},  \frac{6-\lambda ^2}{2 \lambda }<\kappa
   \leq \kappa_+\right\}$, or  $ \left\{\sqrt{2}\leq \lambda <\sqrt{\frac{13}{6}},  \lambda <\kappa \leq \kappa_+\right\}$, or $ \left\{\lambda \leq
   -\sqrt{\frac{13}{6}},  \kappa <\lambda \right\}$, or    $ \left\{-\sqrt{\frac{13}{6}}<\lambda <0,  \kappa <\kappa_-\right\}$, or  $
   \left\{0<\lambda <\sqrt{\frac{13}{6}},  \kappa > \kappa_+\right\}$, or $\left\{\lambda \geq \sqrt{\frac{13}{6}},  \kappa >\lambda
   \right\}$, where we have defined $\kappa_{\pm}=\frac{27-8 \lambda ^2}{16 \lambda }\pm\frac{1}{16} \sqrt{3}
   \sqrt{-\frac{16 \lambda ^2-243}{\lambda ^2}}$. They can be a saddle otherwise. Therefore, these are also hyperbolic
inflationary solutions which can be a future attractor.

Points $B_{4}^{\pm}$ exist for $\lambda^2>2$ and describe
Milne-like solutions, $w_{eff}\left(  B_{4}^{\pm}\right)  =-\frac{1}{3}$ and
$q\left(  B_{4}^{\pm}\right)  =0$ with $\eta\left(  B_{4}^{\pm}\right)
=\pm \sqrt{\frac{\lambda^2}{ 2\left(  \lambda^{2}-1\right)  }}$. The eigenvalues
are $e_{1}\left(  B_{4}^{\pm}\right)  = \pm\sqrt{\frac{2}{\lambda^{2}-1}}\left(
\lambda-\kappa\right)  $, $e_{2}\left(  B_{4}^{\pm}\right)  =\pm\frac
{\lambda+\sqrt{\left(  8-3\lambda^{2}\right)  }}{\sqrt{2\left(  \lambda
^{2}-1\right)  }}$ and $e_{3}\left(  B_{4}^{\pm}\right)  =\pm\frac
{\lambda-\sqrt{\left(  8-3\lambda^{2}\right)  }}{\sqrt{2\left(  \lambda
^{2}-1\right)  }}$. When hyperbolic, $B_4^+$ (respectively $B_4^-$) can be a sink (respectively a source) for $\left\{-2 \sqrt{\frac{2}{3}}<\lambda <-\sqrt{2}, \kappa >\lambda \right\}$, or $\left\{\lambda \leq
-2 \sqrt{\frac{2}{3}}, \kappa >\lambda \right\}$. $B_4^+$ (respectively $B_4^-$) can be a source (respectively a sink) for $\left\{\sqrt{2}<\lambda <2 \sqrt{\frac{2}{3}}, \kappa <\lambda \right\}$, or $\left\{\lambda \geq 2\sqrt{\frac{2}{3}}, \kappa <\lambda \right\}$. They are saddle otherwise. 

Finally,
$B_{5}$ is the vacuum solution  with eigenvalues
$e_{1}\left(  B_{5}\right)  =-\sqrt{2}$, $e_{2}\left(  B_{5}\right)  =-\sqrt{2}$
and $e_{3}\left(  B_{5}\right)  = \sqrt{2}$ that is 
always a saddle.

\section{Hamiltonian analysis and analytic solution}

We proceed our analysis with the construction of analytic solutions for the
field equations. Indeed, the field equations form a Hamiltonian system with
Hamiltonian function%
\begin{align}
& \mathcal{H}=-\frac{p_{a}^{2}}{12a^{2}}+\frac{1}{2a^{4}}\left(  p_{\phi}%
^{2}+e^{-2\kappa\phi}p_{\psi}^{2}\right)  +a^{2}V\left(  \phi\right)  -K,
\label{sp.27}%
\\
& \text{where}\; p_{a}=-6a^{2}a^{\prime}~,~p_{\phi}=a^{4}\phi^{\prime}\text{ and }p_{\psi
}=a^{4}e^{2\kappa\phi}\phi^{\prime}, 
\end{align}
and $a^{\prime}=\frac{da}{d\xi}$ such that $d\xi=adt$.

For the exponential potential $V\left(  \phi\right)  =V_{0}e^{\lambda\phi}$
and for the free parameters $\left(  \kappa,\lambda\right)  =\left(
-\frac{\sqrt{6}}{3},-\frac{\sqrt{6}}{3}\right)  $, the field equations admits
the conservation law%
\begin{equation}
I_{0}=\left(  \sqrt{6}p_{a}+ap_{\phi}\right)  ae^{\frac{\sqrt{6}}{3}\phi}.
\end{equation}
Moreover, an additional conservation law is the
\begin{equation}
I_{1}=a^{4}e^{2\frac{\sqrt{6}}{3}\phi}p_{\psi}.
\end{equation}

Consequently, the conservation laws $\left\{  \mathcal{H},I_{0},I_{1}\right\}
$ are independent and in involution, which means that the field equations form
a Liouville integrable dynamical system.

We follow the procedure presented earlier in \cite{sco01} and we consider the
new variables%
\begin{equation}
a=\left(  \frac{8}{3}\left(  xz-y^{2}\right)  \right)  ^{\frac{1}{4}}%
~,~\phi=\sqrt{\frac{3}{2}}\ln\left(  \sqrt{\frac{2}{3}}\frac{\sqrt{xz-y^{2}}%
}{x}\right),~z=\frac{y}{x},%
\end{equation}
such that the Hamiltonian function to become%
\begin{equation}
\mathcal{H}=\frac{1}{4}p_{y}-p_{x}p_{z}+V_{0}x-K
\end{equation}
and the conservation laws%
\begin{equation}
p_{x}=\Phi_{x}~,~p_{y}=\Phi_{y}.
\end{equation}

Hence, the analytic solution of the field equations is
\begin{equation}
x=x_{1}\xi+x_{0}~,~y=y_{1}\xi+y_{0}~,
\end{equation}
and%
\begin{equation}
z=\frac{\xi^{2}}{2}V_{0}+z_{1}\xi+z_{0}\text{~}.
\end{equation}

Finally, the constraint equation gives%
\begin{equation}
K=y_{1}^{2}-z_{1}x_{1}+x_{0}V_{0}.
\end{equation}
Thus for~$y_{1}^{2}-z_{1}x_{1}+x_{0}V_{0}>0$ the exact solution is valid for a
closed universe while when $y_{1}^{2}-z_{1}x_{1}+x_{0}V_{0}<0$ the exact
solution describes an open universe, while the solution is real in the
original variables when\ $y_{1}^{2}-z_{1}x_{1}<0$. \ At this point it is
important to mention that the same solution is recovered under the change of
variable $\phi\rightarrow-\phi$ if initially we assume $\left(  \kappa
,\lambda\right)  =\left(  \frac{\sqrt{6}}{3},\frac{\sqrt{6}}{3}\right)  $.

Hence, the scale factor for large values of $\xi$, is determined $a\left(
\xi\right)  \simeq\xi^{\frac{1}{2}},$that is, the Hubble function is
$H=\frac{1}{a^{2}}\frac{da}{d\xi}=\frac{1}{\xi^{\frac{3}{2}}}$, that
is, $\left(  H\left(  a\right)  \right)  ^{2}=\frac{1}{4}a^{-6}$, and the
deceleration parameter is derived to be $q\left(  \xi\right)  =2$. That is
nothing else than the asymptotic solution described by points $A_{1}^{\pm}%
\,$\ and $B_{1}^{\pm}$. This means that the results derived from the analysis
of the asymptotic are confirmed by the analytical solution.

\section{Conclusions}

\label{sec4}

In this study, we investigated the dynamics of hyperbolic inflation in the
presence of curvature in a homogeneous and isotropic universe. The question we
wanted to answer is if hyperbolic inflation solves the flatness problem.
Indeed, in an FLRW background space with nonzero curvature, we found as
stationary points the exact solutions which describe the hyperbolic inflation. We focused on the expanding regime and
in the case of a closed ($K=+1$) universe, we found two types of attractors. Say, in the presence of positive spatial curvature we have the solution dominated by quintessence 
$A_{2}$ (respectively $\bar{A}_2$) that is a sink (respectively a source)   when $\lambda^{2}<2$ and $\lambda\left(2\kappa
+\lambda\right)  <6$. Similar conditions were found in the quintom context in \cite{Tot:2022dpr} (under the parameter re-scaling $2\kappa\mapsto \kappa$, $A_2$ corresponds to $\tilde{E}$,  $\bar{A}_2$ corresponds to $\tilde{F}$). 
Moreover, the point $A_3^{\pm}$, that exist when $2\kappa+\lambda\neq0$ and
$\lambda^{2}+2\kappa\lambda-6>0$ and correspond to inflationary power law solutions can be attractors.  $A_3^{\pm}$ (respectively  $\bar{A}_3^{\pm}$) are related to the  point $G$ (respectively $H$) studied in quintom context in \cite{Tot:2022dpr}.

On the other hand, in the case of an open ($K=-1$)  universe, we find, as before, a solution dominated by quintessence 
$B_{2}$ (respectively $\bar{B}_2$), that is a sink (respectively a source)   when $\lambda^{2}<2$ and $\lambda\left(2\kappa
+\lambda\right)  <6$, and  the point $B_3^{\pm}$, that exist when $2\kappa+\lambda\neq0$ and
$\lambda^{2}+2\kappa\lambda-6>0$ and correspond to inflationary power law solutions which can be attractors. Additionally we have the Milne-like (late- and early time-) attractor solutions. Say, 
$B_4^+$ (respectively $B_4^-$) can be a sink (respectively a source) for $\left\{-2 \sqrt{\frac{2}{3}}<\lambda <-\sqrt{2}, \kappa >\lambda \right\}$, or $\left\{\lambda \leq
-2 \sqrt{\frac{2}{3}}, \kappa >\lambda \right\}$. $B_4^+$ (respectively $B_4^-$) can be a source (respectively a sink) for $\left\{\sqrt{2}<\lambda <2 \sqrt{\frac{2}{3}}, \kappa <\lambda \right\}$, or $\left\{\lambda \geq 2\sqrt{\frac{2}{3}}, \kappa <\lambda \right\}$.

From the above analysis, it is clear that in a universe with initial conditions
of nonzero curvature, the hyperbolic inflationary solution exists. We have proved that two hyperbolic inflationary stages are stable solutions that can solve the flatness problem and describe acceleration for both open and closed models, and additionally a Milne-like solution for the open model. We also investigate the contracting branch obtaining mirror solutions with the opposite dynamical behaviours.  That result was also found to be valid with the
derivation of the analytic solution for the problem.

For a more general scalar field potential, it was found that this cosmological
model can be seen as a unified dark energy model \cite{ancqg} because it
provides various epochs of the cosmological evolution. We note that a similar
result will follow in the presence of curvature.

\section*{Acknowledgments} This work is based on the research supported in part by the
National Research Foundation of South Africa (Grant Number 131604).
Additionally, this research is funded by Vicerrector\'{\i}a de
Investigaci\'{o}n y Desarrollo Tecnol\'{o}gico at Universidad Cat\'{o}lica del
Norte. The authors thank Dr. P. Christodoulidis for various comments.

\end{document}